\documentclass[aps,prb,preprintnumbers,english,footinbib,floatfix,reprint,twocolumn,showpacs,amsmath,amssymb,superscriptaddress,longbibliography,10pt,tightenlines]{revtex4-2}
\usepackage{graphicx, epsf,color}

\newcommand{\br}{{\bf r}}
\newcommand{\bu}{{\bf u}}
\newcommand{\ff}{{\mathbf{f}}}

\newcommand{\expv}[1]{\big\langle #1 \big\rangle}
\newcommand{\Expv}[1]{\Big\langle #1 \Big\rangle}

\usepackage{subcaption}
\usepackage{ragged2e}
\DeclareCaptionJustification{justified}{\justifying}
\usepackage{physics}
\usepackage{hyperref}
\hypersetup{
    colorlinks=true,
    linkcolor=blue,
    filecolor=magenta,      
    urlcolor=cyan,
    }

\begin{document}
\title{Sine--Gordon model at finite temperature: the method of random surfaces}
\author{M. T\'oth}
\affiliation{Department of Theoretical Physics, Institute of
Physics, Budapest University of Technology and
Economics, M{\H u}egyetem rkp. 3., H-1111 Budapest,
Hungary}
\affiliation{BME-MTA Statistical Field Theory ’Lend\"ulet’ Research
Group, Budapest University of Technology and
Economics, M{\H u}egyetem rkp. 3., H-1111 Budapest,
Hungary}
\author{J. H. Pixley}
\affiliation{Department of Physics and Astronomy, Center for Materials Theory, Rutgers University, Piscataway, New Jersey 08854, USA}
\affiliation{Center for Computational Quantum Physics, Flatiron Institute, 162 5th Avenue, New York, NY 10010}
\author{D. Sz\'asz-Schagrin}
\affiliation{Department of Theoretical Physics, Institute of
Physics, Budapest University of Technology and
Economics, M{\H u}egyetem rkp. 3., H-1111 Budapest,
Hungary}
\affiliation{BME-MTA Statistical Field Theory ’Lend\"ulet’ Research
Group, Budapest University of Technology and
Economics, M{\H u}egyetem rkp. 3., H-1111 Budapest,
Hungary}
\author{G Tak\'acs}
\affiliation{Department of Theoretical Physics, Institute of
Physics, Budapest University of Technology and
Economics, M{\H u}egyetem rkp. 3., H-1111 Budapest,
Hungary}
\affiliation{BME-MTA Statistical Field Theory ’Lend\"ulet’ Research
Group, Budapest University of Technology and
Economics, M{\H u}egyetem rkp. 3., H-1111 Budapest,
Hungary}
\affiliation{MTA-BME Quantum Dynamics and Correlations
Research Group, Budapest University of Technology and
Economics, M{\H u}egyetem rkp. 3., H-1111 Budapest,
Hungary}
\author{M. Kormos}
\affiliation{Department of Theoretical Physics, Institute of
Physics, Budapest University of Technology and
Economics, M{\H u}egyetem rkp. 3., H-1111 Budapest,
Hungary}
\affiliation{BME-MTA Statistical Field Theory ’Lend\"ulet’ Research
Group, Budapest University of Technology and
Economics, M{\H u}egyetem rkp. 3., H-1111 Budapest,
Hungary}
\affiliation{MTA-BME Quantum Dynamics and Correlations
Research Group, Budapest University of Technology and
Economics, M{\H u}egyetem rkp. 3., H-1111 Budapest,
Hungary}

\date{15th August 2024}

\begin{abstract}
We study the sine--Gordon quantum field theory at finite temperature by generalizing the method of random surfaces to compute the {free energy} and one-point functions of exponential operators non-perturbatively. Focusing on the gapped phase of the sine--Gordon model, we demonstrate the method's accuracy by comparing our results to {the predictions of other methods} and to exact results in the thermodynamic limit. We find excellent agreement between the method of random surfaces and other approaches when {the temperature is not too small with respect to the mass gap}. Extending the method to more general problems in strongly interacting one-dimensional quantum systems is discussed.
\end{abstract}

\maketitle

\section{Introduction}

Sine--Gordon (sG) quantum field theory is the low-energy (or equivalently, long-distance) description of a large spectrum of gapped one-dimensional (1D) systems \cite{Giamarchi:743140}, with numerous applications ranging from quasi-1D antiferromagnets, carbon nanotubes and organic conductors \cite{Controzzi2001,2005ffsc.book..684E},
trapped ultra-cold atoms \cite{2007PhRvB..75q4511G,2010PhRvL.105s0403C,2010Natur.466..597H,2017Natur.545..323S,PRXQuantum.4.030308}, to quantum circuits \cite{2021NuPhB.96815445R} and coupled spin chains \cite{Wybo2022}. It is also a paradigmatic example of quantum field theory with highly nontrivial non-perturbative dynamics and a strong-weak coupling duality \cite{1975PhRvD..11.2088C,1975PhRvD..11.3026M}. Due to its integrability, many exact results are available, including its exact $S$-matrix \cite{1977CMaPh..55..183Z,ZAMOLODCHIKOV1979253}, 
its thermodynamics via the Thermodynamic Bethe Ansatz \cite{2024ScPP...16..145N} and the nonlinear integral equation 
\cite{1991JPhA...24.3111K,1995NuPhB.438..413D},
its form factors \cite{1992ASMP...14....1S} and exact zero-temperature expectation values of local fields \cite{Lukyanov1996}. However, determining finite temperature expectation values and, more generally, correlation functions presents serious challenges.

Correlation functions are among the most fundamental quantities characterizing spatial correlations and dynamics and are also measurable in experiments. Besides their use in reconstructing physical quantities such as susceptibility or conductivity, cold atom experiments can directly access sG correlation functions using atom wave interferometry \cite{2017Natur.545..323S}. Despite the integrability of the sine--Gordon model, the available theoretical approaches face severe limitations. Perturbative expansions only work for short distances / high temperatures. For zero temperature correlators, a long-distance expansion can be written in terms of the exact form factors \cite{2005ffsc.book..684E}, which also enables the computation of the low-temperature behavior of expectation values of local fields \cite{2014JHEP...03..026B}. However, extending the form factor expansion to finite temperature is much more limited in scope \cite{2009JSMTE..09..018E,2010JSMTE..11..012P}. 

An alternative is to consider approaches which are independent of integrability. One possible approach is discretizing the fields, putting the theory on a space-(imaginary) time lattice, and using standard Monte Carlo techniques. The classical 2D sine--Gordon model was studied this way in connection with the BKT roughening transition \cite{1994PhyA..211..255H,2000PhRvE..62.3219S,Flamino:2018jmo}. However, from the point of view of the 1D quantum sine--Gordon theory, these studies were carried out at zero temperature (infinite 2D geometry).

Another possibility is provided by Hamiltonian truncation \cite{Yurov:1989yu}, which was extended to the sG theory some time ago \cite{1998PhLB..430..264F}; for recent reviews see Refs.\ \cite{2018RPPh...81d6002J,2022CoPhC.27708376H}. It was recently utilized to evaluate the time evolution of correlation functions both in and out of equilibrium \cite{2018PhRvL.121k0402K,2020JHEP...07..224K}, and provides a fully non-perturbative approach to quantum dynamics. Still, it is limited by the truncation to low energy densities and by finite-size effects that appear due to the finite volume setting \footnote{We remark that an alternative formulation of the truncated Hamiltonian method using light-cone quantization works directly in infinite volume \cite{2016JHEP...07..140K,2020arXiv200513544A}.}. 

Here, we consider an alternative numerical approach to sine--Gordon quantum field theory that relies on reformulating the problem in terms of random surfaces, which was previously used to compute full distribution functions of interference amplitudes \cite{2008PhRvA..77f3606I}. The method of random surfaces (MRS) is a type of Monte Carlo method, with the main advantage given by its simplicity and speed, with the necessary computing resources scaling polynomially with the number of modes retained in the computation. It is also free of the well-known sign problem and can access multi-point functions (including finite-temperature ones), which is a great advantage given their experimental availability \cite{2017Natur.545..323S}. In addition, it is also possible to implement spatially inhomogeneous couplings, which are generally present in experimental realizations due to the traps used to localize the atomic clouds. Previous applications of the method include experimental analysis of variations of interference fringe contrast for pairs of independently created one-dimensional Bose condensates \cite{2008NatPh...4..489H}, and dissipative Landau--Zener transitions \cite{2010PhRvA..82c2118O,2013PhRvB..87a4305O}.

The present paper is devoted to developing the method for the sine--Gordon model and demonstrating its validity and efficiency, with results for the free energy and operator expectation values validated by using the non-linear integral equation (NLIE) as well as the minisuperspace extended truncated Hamiltonian approach (MSTHA) previously developed to study non-equilibrium dynamics in the model \cite{2024PhRvB.109a4308S}. Application of the MRS approach to more general correlation functions is relegated to a follow-up paper \cite{mrs_correlators}.

The outline of the present work is as follows. Section \ref{sec:model_method} introduces the model and the method of random surfaces, the latter in two stages: first for the partition function and then for the expectation values of vertex operators. Section \ref{sec:results} presents our numerical results and their validation using the NLIE and the MSTHA, while we give our conclusions and outlook in Section \ref{sec:conclusions}. To keep the main argument uninterrupted, some details were relegated to the appendices including technical aspects of the method of random surfaces, a short summary of the NLIE and some additional numerical results.

\section{Model and Method}\label{sec:model_method}

In this section, we introduce the sine--Gordon model and provide a detailed derivation of the partition function and one-point functions using the MRS approach.

\subsection{The sine--Gordon model}
The sine--Gordon model is a $1+1$-dimensional relativistic quantum field theory, which at finite temperature $T$ is defined by the Euclidean action 
\begin{align}
    S_\mathrm{sG} &= S_0 + S_I\,, \nonumber\\
    S_0&=\int_{-\infty}^\infty dx\int_0^R d\tau\frac{1}{2}
    \left[(\partial_{\tau} \phi)^2 +  (\partial_{x} \phi)^2\right]\,, \nonumber\\
    S_I&=- \lambda_0 \int_{-\infty}^\infty dx\int_0^R d\tau \cos(\beta\phi) \,.
\end{align}
Here $R=1/T$ is the inverse temperature, $0\le\tau\le R$ is the imaginary time, and $\phi(x, \tau)$ is a real scalar field satisfying periodic boundary conditions in the $\tau$ direction: $\phi(x,\tau+R)=\phi(x,\tau)$. We work with units such that both the Boltzmann constant and the speed of light are one ($k_\text{B}=1$, $c=1$).

The model can be viewed as a perturbed conformal field theory where the free boson CFT with action $S_0$ is perturbed by the $\cos(\beta\phi)$ operator, which is relevant for $0<\beta^2<8\pi$. This work will focus on the so-called attractive regime $0<\beta^2<4\pi$. 

The partition function of the model at an inverse temperature $R$ can be written as a coherent state bosonic path integral $Z=Z_0Z_I$, where $Z_0=\int D[\phi]\, e^{-S_0}$ is the partition function of the free boson, and
\begin{align}
    Z_I = \displaystyle\frac{\int D[\phi]\, e^{-S_0-S_I}}{\int D[\phi]\, e^{-S_0}} 
    \equiv \langle e^{-S_I} \rangle_0 \,,
    \label{eqn:avg0}
\end{align}
where we have introduced the average over free bosonic fields $\langle \dots \rangle_0 = \int D[\phi]\, (\dots)e^{-S_0}/Z_0$.

Using this representation, the partition function can be evaluated by the MRS introduced in Ref.\ \cite{2008PhRvA..77f3606I}. Here, we develop the MRS further to compute the free energy $F=-R\ln Z$ and extend it to the one point functions $\langle e^{i\alpha \phi} \rangle$. In both cases, we ascertain the accuracy and speed of the MRS when compared to other known available techniques to study the sG model. 

\subsection{Method of random surfaces: partition function}\label{subsec:mrs_partition_function}

The MRS can be derived by considering a perturbative expansion of $Z$ in terms of $\lambda_0$
\begin{align}
\label{eq:Zexpansion}
Z_I &= \sum_{n=0}^{\infty} \frac{1}{n!}\left( \frac{\lambda_0}{2} \right)^{n}\tilde Z_{n}\,,\\
\tilde Z_n &= \prod_{j=1}^n \int dx_j \int_0^R d\tau_j \Big \langle \prod_{j=1}^n \left( e^{i\beta\hat\phi(\br_j)} + e^{-i\beta\hat\phi(\br_j)} \right) \Big \rangle_0\,,
\end{align}
where $\br_i = (x_i, \tau_i)$ and the subscript 0 refers to the expectation value in the unperturbed free boson theory as defined in Eq.\ \eqref{eqn:avg0}. Expanding the product leads to the sum of $n$-point functions of the operators $e^{\pm i\beta\hat\phi(\br_i)}$. The multi-point functions appearing in the integrand can be calculated by carrying out the Gaussian path integral, which leads to the multiplicative form of Wick's theorem:
\begin{align}   
    \Big \langle \prod_{j=1}^n  e^{i\alpha_j \hat\phi(\br_j)} \Big\rangle_{0}
    = 
    e^{-\sum\limits_{j<k}^n \alpha_j \alpha_k \tilde G(\br_j, \br_k)-\sum\limits_j^n \alpha_j^2 \tilde G(\br_j,\br_j)}\,,
    \label{eq:expvalpartfunc}
\end{align}
if the so-called neutrality condition $\sum_j\alpha_j=0$ is satisfied and zero otherwise \footnote{This is related to the fact that the Laplace operator on the cylinder has a zero eigenvalue. It also reflects the $\phi\to\phi+$const. symmetry of the action.}. Here 
\begin{align}
\label{eq:Greens}
\tilde G(\br_j, \br_k) 
= -\frac1{4\pi} \log \left|\frac1{\pi}\sinh\left(\frac{\pi}{R}(x_{jk}+i\tau_{jk}+a)\right)\right|^2
\end{align}
is the Green's function, i.e.\ a solution of the Poisson equation on an infinite cylinder of circumference $R$. Here we introduced a regulator length $a$, and $(x_{jk},\tau_{jk})=(x_j-x_k,\tau_j-\tau_k)$ so the Green's function is translational invariant, moreover,
it is periodic in the $\tau$ direction with period $R$. 

Plugging  Eq.\ \eqref{eq:Greens} into Eq.\ \eqref{eq:expvalpartfunc} we obtain 

\begin{widetext}
\begin{align}
\label{eq:multipoint}
\left \langle \prod_{j=1}^n \left( e^{i\alpha_j \hat\phi(\br_j)} \right) \right \rangle_0
= 
a^{\frac{\sum_j\alpha_j^2}{4\pi}}\left(\frac{1}R\right)^{\frac{(\sum_j\alpha_j)^2}{4\pi}} \prod_{j<k}\left|\frac{R}\pi \sinh\left(\frac{\pi}{R}(x_{ij}+i\tau_{ij}+a)\right)\right|^{\frac{\alpha_j\alpha_k}{2\pi}}\,.
\end{align}
\end{widetext}

The neutrality condition ensures the result is finite in the $R\to\infty$ (zero temperature) limit. It is useful to note that $\tilde G(\br_j,\br_k)$ can be shifted by an arbitrary constant without changing the result in Eq.~\eqref{eq:multipoint}, which we will exploit shortly. 

Substituting Eq.\ \eqref{eq:multipoint} into Eq.\ \eqref{eq:Zexpansion} and collecting the combinatorial factors we arrive at
\begin{subequations}
\label{eq:Coulomb}    
\begin{align}
Z_I(\lambda,\Delta,R) &= \sum_{m=0}^{\infty} \frac{1}{(m!)^2}\left( \frac{\lambda}{2R^\Delta} \right)^{2m}Z_{2m}(\Delta,R)\,,\label{eq:ZIexp2}\\
Z_{2m}(\Delta,R)&=\prod_{j=1}^{2m}\int d\br_j \exp\left[-\Delta\sum_{j<k}^{2m}\epsilon_j\epsilon_k G(\br_j,\br_k)\right], \label{eq:Z2m}
\end{align}
\end{subequations}
where only the $n=2m$ even terms survive due to the neutrality condition. The $R$-dependence is only through $G(\br_j,\br_k)=4\pi \tilde G(\br_j,\br_k)$, $\Delta = \beta^2/4\pi$ is the scaling dimension of the operators $e^{\pm i\beta\hat\phi(\br)}$, and $\epsilon_j=\pm1$, $m$ of them being $+1$ and $m$ being $-1$. The explicit $a$-power in the correlation function \eqref{eq:multipoint} has been absorbed into 
a multiplicative renormalization of the coupling constant, 
\begin{align}
\label{eq:lambda}
    \lambda=a^\frac{\beta^2}{4\pi} \lambda_0\,.
\end{align}
If we define the so-called vertex operators as \footnote{This is equivalent to normal ordering the exponential operator with respect to the modes of the massless bosonic field governed by the free action $S_0$.} 
\begin{align}
\label{eq:vertex}
    \hat V_\alpha(\br) = a^{-\frac{\alpha^2}{4\pi}} e^{i\alpha\hat\phi(\br)}\,,
\end{align}
then the perturbation can be written as
\begin{align}
\label{eq:vertexpert}
    \lambda_0 \cos \beta\hat\phi(\br) = \frac\lambda2 \left(\hat V_\beta(\br) + \hat V_{-\beta}(\br)  \right)\,.
\end{align}
From Eq. \eqref{eq:multipoint} we can read off the two-point function of vertex operators,
\begin{align}
\label{eq:vertexcorr}
    \expval{V_\alpha(\br_1)V_{-\alpha}(\br_2)} = \left|\frac{R}\pi \sinh\left(\frac{\pi}{R}(x_{12}+i\tau_{12}+a)\right)\right|^{-\frac{\alpha^2}{2\pi}}
\end{align}
which for $|\br_{12}|\to0$ behaves as $\expval{V_\alpha(\br_1)V_{-\alpha}(\br_2)} \sim 
a^{-\alpha^2/(2\pi)}$ 
(or without the regulator $a$,  $\expval{V_\alpha(\br_1)V_{-\alpha}(\br_2)}\sim 
|\br_{12}|^{-\alpha^2/(2\pi)}$), obeying the standard conformal normalization of scaling operators. 

In the attractive regime $\beta^2<4\pi$ ($\Delta<1$), the multiple integrals in Eq.\ \eqref{eq:Z2m} are finite as $a\to0$ so we are allowed to drop $a$ in the expression of $G(\br_j,\br_k)$. 

The series in Eqs.\ \eqref{eq:Coulomb} can be interpreted as the grand canonical partition function of a classical 2D system of particles with electric charge $\epsilon_j$ that interact via the 2D Coulomb potential $G(\br_j,\br_k)$. In this picture, $\Delta$ is related to the inverse temperature, while $\lambda R^{-\Delta}$ is related to the fugacity of the classical system. The Coulomb gas partition function in Eq.~\eqref{eq:Coulomb} appears in the field theoretical description of the 2D $XY$-model, 2D superfluids, and thin superconducting films \cite{BKT_book}.

The first step in the MRS approach is to write $G(\br_1,\br_2)$ in a projector decomposition form (i.e. diagonalize with Fourier series)
\begin{align}
\label{eq:Gdecomp}
    G(\br_1,\br_2) = \sum_\ff G_\ff \psi_\ff(\br_1)\psi_{\ff}(\br_2)\,.
\end{align}
This can be achieved by expanding the left-hand side in a double Fourier series. To obtain a discrete series, we restrict integration over the $x_j$ variables to the interval $[-L/2,L/2]$ and extend the function $G(\br_{12},0)$ on $x_{12}\in [-L,L]$ periodically in the $x$ direction with period $2L$ {(see Appendix \ref{app:modes} for details)}. 
As a result, we integrate over a finite cylinder but using the Green's function of the infinite cylinder. This introduces a finite-size effect that must be addressed in the calculation.
Since $G$ is a real symmetric function, it can be expanded in a double cosine series, so
\begin{align}
G(\br_1,\br_2)= \sum_{m,n}^{\infty}A_{mn}\cos(\pi mx_{12}/L)\cos(2\pi n \tau_{12}/R)
\end{align}
which, upon using trigonometric subtraction identities, leads to the form of Eq.\ \eqref{eq:Gdecomp}
where $\ff=(m,n,\ell)$ is a multi-index with integers $m,n>0$, $1\le \ell\le4$, and the eigenfunctions for $m\neq0,n\neq0$ are 
\begin{subequations}
\begin{align}
\label{eq:psi}
\psi_{m,n,1}(\br)&=\cos(\pi m x/L)\cos(2\pi n\tau/R)\,,\\
\psi_{m,n,2}(\br)&=\cos(\pi m x/L)\sin(2\pi n\tau/R)\,,\\
\psi_{m,n,3}(\br)&=\sin(\pi m x/L)\cos(2\pi n\tau/R)\,,\\
\psi_{m,n,4}(\br)&=\sin(\pi m x/L)\sin(2\pi n\tau/R)\,.
\end{align}
\end{subequations}
If $m=0,n\neq0$ or $n=0,m\neq0$, there are only two functions (sin and cos), and for $m=n=0$ there is only one, the $\psi_{0,0,1}$ constant term.

The coefficients $G_{(m,n,\ell)}=A_{mn}$ can be obtained by numerical integration. 
For $L/R\gtrsim1$, they turn out to be very close to the following simple expressions (cf. Appendix \ref{app:modes}):
\begin{subequations}
\label{eq:Amn}
\begin{align}
  A_{00} = 2\log(2\pi) - \pi L/R\,,
\end{align}

\begin{align}
  A_{m0} = \begin{cases}
    \frac{8L}{\pi R} \frac1{m^2} & \quad\text{m odd,} \\
    0 & \quad\text{m even, } m\neq0\,,
\end{cases}
\end{align}
and
\begin{align}
    A_{mn} = \frac{1}{1 + \delta_{m,0}} \frac{8\pi}{LR} \frac{1}{\left(\frac{\pi m}{L} \right)^2 + \left(\frac{2\pi n}{R} \right)^2}\,,\qquad n\neq0\,.
\end{align}
\end{subequations}
The number of Fourier modes is truncated in a numerical calculation to a finite value $m_\text{max}$, which naturally introduces an ultraviolet cutoff. From now on, we assume that the number of modes is finite and vary $m_\text{max}$ to ensure that the results have converged at a given $L/R$ ratio and dimensionless coupling constant $\lambda R^{-\Delta}$. 

Using the decomposition in Eq.\ \eqref{eq:Gdecomp} in Eq.\ \eqref{eq:Z2m}, we can write $Z_{2m}$ as
\begin{widetext}
\begin{align}
    Z_{2m} =  \prod_{j=1}^{2m}\int d\br_j  
    \exp \left\{ -\frac{\Delta}{2}\sum_\ff G_\ff \left( \sum_{j=1}^{2m} \epsilon_j\psi_\ff(\br_j)\right)^2 \right\}
    \exp\left\{ \frac{\Delta}{2}\sum_\ff G_\ff \sum_{j=1}^{2m} \psi_\ff(\br_j)^2  \right\}.
\end{align}
\end{widetext}
The last term factorizes into a $2m$-fold product in terms of the integration variables. However, due to the specific trigonometric form of the $\psi_\ff$ functions in Eq.\ \eqref{eq:psi}, it is a constant equal to $\exp(-m\Delta\sum_{m,n}A_{mn})$.
To factorize the first term, we observe that all $G_\ff$ are positive except $G_{0,0,1}$ for which the whole exponent is zero due to the $\sum_j \epsilon_j=0$ neutrality condition. To avoid numerical instabilities, we set $G_{0,0,1}$ to zero, which amounts to a constant shift of $G(\br_1,\br_2)$, which can be accounted for at the end of the calculation.
We can thus apply a Hubbard--Stratonovich decoupling using the identity
\begin{align}
    e^{-\frac{1}{2}x^2} = \sqrt{\frac{1}{2\pi }} \int_{-\infty}^\infty e^{-\frac{y^2}{2} \pm ixy} dy
\end{align}
and introducing a Gaussian field $t_\ff$ for each mode, which yields
\begin{widetext}
\begin{align}
    \prod_\ff \exp \left\{ -\frac{\Delta}{2}G_\ff\left( \sum_{j=1}^{2m} \epsilon_\ff\psi_\ff (\br_j) \right)^2\right\}=
    \prod_\ff \int_{-\infty}^\infty \frac{dt_\ff}{\sqrt{2\pi}} e^{-\frac{t_\ff^2}{2}} \exp \left\{ i t_\ff\sqrt{\Delta G_\ff}\sum_{j=1}^m \epsilon_j\psi_\ff (\br_j) \right\}.
\end{align}
\end{widetext}
Swapping the integrals over the $t_\ff$ with the space-time integrals, the latter factorizes to a product of $m+m$ identical integrals. This leads to
\begin{align}
    Z_{2m} = \int D[t_\ff]\,\big[g(\{t_\ff\})g(\{-t_\ff\})\big]^m
    \,,
    \label{eq:z2mwithds}
\end{align}
where $\int D[t_\ff] = \prod_\ff \int_{-\infty}^\infty \frac{dt_\ff}{\sqrt{2\pi}} e^{-t_\ff^2/2}$
and
\begin{align}
    g(\{t_\ff\}) = C\int d\bu \exp\left\{ \sum_\ff^{m_{\mathrm{max}}}  it_\ff\sqrt{\Delta G_\ff}\,\psi_\ff(\bu) 
    \right\}
    \label{eq:gtf}
\end{align}
with a prefactor 
\begin{align}
    C=\exp\left(
    \frac{\Delta}{2}
    \sum_{m,n=1}^{m_{\mathrm{max}}}A_{mn}\right).
\end{align} 
The high-dimensional integral over $\{t_\ff\}$ can be performed in a Monte Carlo fashion by averaging the random $t_\ff$ variables drawn from a Gaussian distribution with zero mean and unit variance. In this picture, the integrand of $g(\{t_\ff\})$ is a random function of two variables, i.e.\ a fluctuating surface, hence the method's name.

Plugging $Z_{2m}$ in Eq. \eqref{eq:z2mwithds} into the expansion \eqref{eq:ZIexp2}, and \emph{swapping the integral with the infinite sum}, the latter can be carried out exactly to yield
\begin{align}    
    Z_I(\lambda) = \int D[t_\ff] \,I_0 \left( \lambda R^{-\Delta} \sqrt{g(\{ t_\ff\})g(\{ -t_\ff\})}\right)\,,
\end{align}
where $I_0(x)$ is the modified Bessel function of the first kind. Remarkably, this sums up the entire perturbative series, \emph{exactly}. Importantly, as demonstrated by the following numerical calculations presented in Sec.~\ref{sec:results}, the MRS result \emph{fully captures all the nonperturbative contributions} as well, which is a feature arising (somewhat miraculously) from the above manipulations.

In the numerical calculations, dimensionful quantities are measured in powers of $R$, i.e.\ one sets $R=1$. We choose an $L/R$ ratio and determine the coefficients $A_{mn}$ up to a maximal mode number $m_\text{max}$. We then draw a set of random $\{t_\ff\}=\{t_{(m,n,\ell)}\}$ from the Gaussian distribution with zero mean and variance equal to unity. We compute the 2D integral \eqref{eq:gtf} for $g(\{t_\ff\})$ by discretizing the mode functions $\psi_\ff$ on a grid. We study the effect of the finite resolution of the grid in Appendix \ref{app:discr}.
$Z_I$ is then obtained by averaging over $N_\text{s}$ different $\{t_\ff\}$ samples. Note that once the $g(\{t_\ff\})$ are computed and stored, the partition function can be easily evaluated for any value of the coupling $\lambda$, showcasing one powerful aspect of the MRS approach. 

In the following, we consider aspect ratios of the two-dimensional system in the range $6\lesssim L/R\lesssim12$; this requires taking on the order of $20\lesssim m_\text{max}\lesssim 60$ Fourier modes and the number of Monte Carlo samples $10^4\lesssim N_\text{s}\lesssim10^6$. A major benefit of the MRS approach to the sG model is that it is simple to implement (the length of the necessary code in C is of the order of 50 lines), understand, and execute. Moreover, being a Monte Carlo technique, it is amenable to parallelization, which trivially reduces the running time.
However, it is far from obvious how the accuracy of the MRS approach depends on the parameters used in it, and therefore, it is imperative to compare against exact results to establish both the validity and power of the technique. As a result, we can understand the convergence of the MRS as we tune the number of Fourier modes, the number of Monte Carlo samples, and the numerical accuracy in the numerical integration of Eq.~\eqref{eq:gtf}. Therefore, the present work serves as a benchmark for the MRS and a stepping stone for the MRS to be further applied to compute properties of models where such comparisons are no longer possible.

\subsection{Method of random surfaces: vertex operator expectation values}\label{subsec:mrs_vertex_operators}

The MRS method can also be used to compute observables that can be measured experimentally, such as expectation values and correlation functions of local operators. One family of such observables is the finite temperature expectation values of the vertex operators $ \hat V_\alpha(\br)$ defined in Eq.\ \eqref{eq:vertex}. From the theoretical perspective, these are truly nonperturbative objects in an interacting quantum field theory, so they are of fundamental interest. From a more pragmatic standpoint,
they are related to physical observables in the various realizations of the sine--Gordon model. For example, in the tunnel-coupled cold atomic condensates system, this expectation value is called the coherence factor and is routinely measured via matter-wave interferometry \cite{Hofferberth2007,Nieuwkerk2018}. Comparing the measured values with theoretical results provides important information about the effective temperature so it can be used for thermometry.

For generic exponent $\alpha$, the expectation value can be written as
\begin{align}
    \expv{ \hat V_\alpha(\br)} = {Z^{-1}Z_0}\expv{a^{-\frac{\alpha^2}{4\pi}} e^{i\alpha\hat\phi(\br)}e^{-S_I(\lambda)}}_0
\end{align}
Expanding the numerator results in
\begin{widetext}
\begin{align}
   \expv{ \hat V_\alpha(\br)}  = \frac{{1}}{Z_I(\lambda)} \sum_{n=0}^\infty \frac{1}{n!}\left(\frac{\lambda_0}2\right)^n \prod_{j=1}^n \int dx_j \int_0^R d\tau_j \,
   a^{-\frac{\alpha^2}{4\pi}}\,\Expv{e^{i\alpha \phi(\br)} \prod_{j=1}^n \left( e^{i\beta\phi(\br_j)} + e^{-i\beta\phi(\br_j)} \right)}_0\,.
\end{align}
\end{widetext}
Due to the neutrality condition, the expectation value in the integral is only non-vanishing when $\alpha$ is an integer multiple of $\beta$, i.e.\  $\alpha = s\beta$ with $s\in\mathbb{Z}$. The non-vanishing neutral terms are those where out of the $n$ terms, we pick the $+$ exponentials from $n_+=(n-s)/2$ and the $-$ exponentials from $n_-=(n+s)/2$, so the parity of $s$ and $n$ must be the same. This gives
\begin{align}
\label{eq:Valphaseries}
  \expv{ \hat V_\alpha(\br)}  =  \frac{1}{Z_I(\lambda)} \sideset{}{'}\sum_{n=0}^\infty \frac{1}{n_+!\,n_-!} \left(\frac{\lambda}{2R^\Delta}\right)^n V_{n}\,,
\end{align}
where the primed sum runs over integers $n$ that have the same parity as $s$, $n_++n_-=n$, $n_+-n_-=s$, and 
\begin{widetext}
\begin{align}
    V_n = \prod_{j=1}^n \int d\br_j 
    \,a^{-\frac{\alpha^2}{4\pi}}\,
    \bigg\langle e^{is\beta\phi(\br)} 
    \prod_{j=1}^{n_+}e^{i\beta\phi(\br_j)} 
    \prod_{j=1}^{n_-}e^{-i\beta\phi(\br_j)} \bigg\rangle_0 
    =
    R^{-\Delta_\alpha}\prod_j \int d\br_j \,
    e^{-\Delta \sum_{j<k}^n \epsilon_j \epsilon_k G(\br_i, \br_j)} e^{-s\Delta \sum_j^n \epsilon_j G(\br,\br_j)}\,,
\label{eq:Vn}
\end{align}
\end{widetext}
where $\Delta_\alpha=\alpha^2/(4\pi)$ and now $n_+$ of the signs are $\epsilon_j=+1$ and $n_-$ of them is $\epsilon_j=-1$. We proceed as above and decouple the integration variables using the Hubbard--Stratonovich transformation. It turns out to be convenient to perform the decoupling in the $G(\br,\br_j)$ terms as well, which leads to
\begin{align}
     V_n =R^{-\Delta_\alpha} \int D[t_\ff] \,g(\{t_\ff\})^{n_+} g(\{-t_\ff\})^{n_-}  
     h_\alpha(\{t_\ff\}, \br)
     \,,
\end{align}
where
\begin{equation}
h_\alpha(\{t_\ff\}, \br) = C_s \exp(\sum_\ff\left[ i t_\ff s\sqrt{\Delta G_\ff} \psi_\ff(\br) \right])
\end{equation}
with $s=\alpha/\beta$ and $C_s=\exp{\Delta s^2/2\,\sum_{m,n}A_{mn}}=C^{s^2}$. Plugging everything back into Eq.\ \eqref{eq:Valphaseries}, the infinite series can again be summed up, yielding
\begin{widetext}
\begin{align}
\label{eq:Valpha2}
    \expv{ \hat V_\alpha(\br)}  
    =\frac{R^{-\Delta_\alpha}}{Z_I(\lambda)} \int D[t_\ff]\, 
    \left(\frac{g(\{-t_\ff\})}{g(\{t_\ff\})}\right)^{\frac\alpha{2\beta}} 
    h_\alpha(\{t_\ff\}, \br)\,
    I_{\left|\frac\alpha\beta\right|}\left( \lambda R^{-\Delta} \sqrt{g(\{t_\ff\})g(\{-t_\ff\})}\right)\,.
\end{align}
\end{widetext}

Compared to the partition function, the new ingredient is $h_\alpha(\{t_\ff\},\br)$, the only term depending on the position $\br$. In the ideal $L\to\infty$ case, the final result should be independent of the position $\br$, but in a numerical calculation, there are deviations near the integration boundaries $x=\pm L/2$. As shown in Appendix \ref{app:posdependence}, these deviations decay exponentially fast with the distance from the boundaries.

It is easy to see  from Eq.\ \eqref{eq:Valpha2} that $\expv{ \hat V_{-\alpha}(\br)}=\expv{ \hat V_\alpha(\br)}^*$, as it should be. However, the result for $\expv{ \hat V_\alpha(\br)} $ is real, so 
\begin{align}
\label{eq:Vreal}
    \expv{ \hat V_\alpha(\br)} = \expv{ \hat V_{-\alpha}(\br)} = 
    a^{-\frac{\alpha^2}{4\pi}} \expv{\cos\alpha\hat\phi(\br)}\,.    
\end{align}
As a non-trivial consistency check, consider the case $\alpha=\beta$ and integrate both sides of Eq.\ \eqref{eq:Valpha2} over $\br$. Using $\int h_\beta(\{t_\ff\}, \br) d\br= g(\{t_\ff\})$
we find
\begin{widetext}
\begin{align}
     \int d\br \expv{ \hat V_\beta(\br)}
     = \frac{R^{-\Delta}}{Z_I(\lambda)} \int DS\, I_1\left( \lambda R^{-\Delta} \sqrt{g(\{t_\ff\}) g(\{-t_\ff\})} \right) \sqrt{g(\{t_\ff\}) g(\{-t_\ff\})}   =\frac{\partial \log(Z)}{\partial \lambda}\,,
\end{align}
\end{widetext}
where in the last step we used $I_0'(x)=I_1(x)$. Since from Eqs.\ \eqref{eq:lambda}, \eqref{eq:vertex}, and \eqref{eq:Vreal} $\lambda_0\expv{\cos\beta\hat\phi(\br)}=\lambda\expv{ \hat V_\beta(\br)}$, the above result is nothing but the  Hellmann--Feynman theorem \cite{hellman1937einfuhrung,1939PhRv...56..340F}, demonstrating the validity of the MRS equations.

\section{Results }\label{sec:results}

This section presents our numerical results obtained with the MRS approach discussed in the previous section. To provide evidence for the strong claim about the nonperturbative nature of the approach and to justify the somewhat dubious steps, such as swapping integrals and summation in the perturbative expansion during the derivations in Subsections \ref{subsec:mrs_partition_function} and \ref{subsec:mrs_vertex_operators}, we compare our predictions with exact nonperturbative results obtained by exploiting the integrability of the sine--Gordon model via the nonlinear integral equation (NLIE) approach, as well as with numerical results from the MSTHA.

\subsection{Free energy}

\begin{figure}[t!]
    \centering
    \includegraphics[width=\columnwidth]{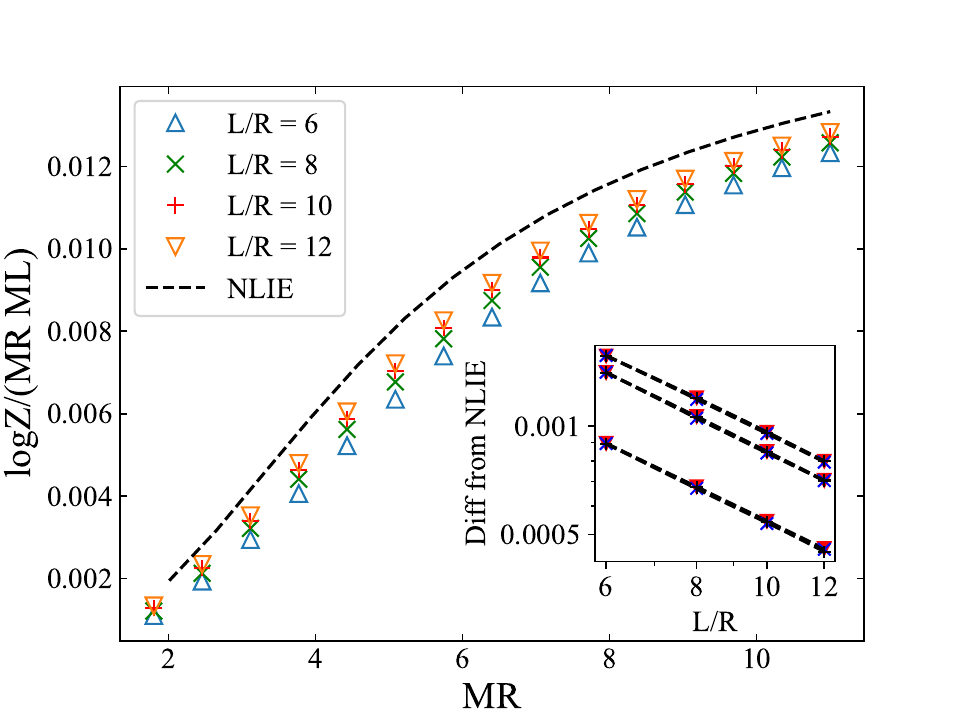}
    \caption{Free energy density at $\beta^2=8\pi/25$ ($\Delta=2/25$) as a function of the dimensionless inverse temperature $MR$ for four different cylinder ratios $L/R=6,8,10,12$ for a fixed maximal mode $m_\text{max}=20$. The inset shows the difference between the NLIE and MRS results for three different inverse temperatures $MR = 4,8,12$ and for different mode cutoffs ($m_\text{max}=20$ in red triangle, 25 in blue crosses, 30 in black plus signs). We find the results are smoothly approaching the thermodynamic limit and have already converged as a function of the number of Fourier modes at $m_\text{max}=20$. The MRS results were obtained using $\sim10^6$ Monte Carlo samples, the statistical errors are smaller than the symbol size.
    }
    \label{fig:fK01}
\end{figure}

Let us start with the free energy density, which in our approach is given by
\begin{align}
  f =  \frac{F}L=-\frac{\log Z}{LR}  = -\frac{\log( Z_IZ_0)}{LR} =-\frac{\log Z_I}{LR}+\frac{\pi}{6R^2}\,,
\end{align}
where the last term is the free energy density of the free massless boson at inverse temperature $R=1/T$. 

The exact free energy can be computed from the NLIE, which we review in Appendix \ref{app:nlie}. It can also be determined using the Thermodynamic Bethe Ansatz, which was derived recently  for general (rational) values of the coupling parameter \cite{2024ScPP...16..145N}
\begin{align}
    \xi=\beta^2/(8\pi-\beta^2)\,.
    \label{eqn:xi}
\end{align}
The exact relation between the (renormalized) perturbing parameter $\lambda$ and the soliton mass is \cite{1995IJMPA..10.1125Z}
\begin{align}
  \lambda = \kappa(\xi) M^{2/(\xi+1)},
\end{align}
where 
\begin{align}
  \kappa(\xi) = \frac{1}{\pi} \frac{\Gamma\left(\frac{\xi}{\xi+1}\right)}{\Gamma\left(\frac{1}{\xi+1}\right)} \left[ \frac{\sqrt{\pi} \Gamma\left( \frac{\xi+1}{2} \right)}{2 \Gamma \left( \frac{\xi}{2} \right)} \right]^{2/(\xi+1)}.
\end{align}
We then obtain the free energy density in the form
\begin{align}
  f(\xi,\Delta,R) = \tilde{f}(MR) + B(\xi)\,,
  \label{eq:free_energy_definition}
\end{align}
where $\tilde f(MR)$ is the free energy density calculated from the NLIE/TBA, and $B(\lambda)$ is the so-called bulk energy density
\begin{align}
  B(\xi) = -\frac{M^2}{4}\tan\frac{\pi \xi}{2}
\end{align}
which accounts for the different normalization of the free energy. Alternatively, $f/T$ can also be interpreted as the (zero-temperature) ground state energy in a finite volume $R=1/T$.

In the MRS calculation, we can conveniently tune the dimensionless inverse temperature $MR$ through $M$ by changing $\lambda$. The resulting free energy density is plotted in Fig. \ref{fig:fK01} for four different $L/R$ ratios and the NLIE result (dashed line). {The deviation is a finite size effect which decreases with increasing $L/R$. This is shown in the inset, where the difference between our results and the NLIE value is shown as a function of $L/R$ on a log-log scale at three different $MR$ values. The symbols in the inset represent three different mode cutoffs, demonstrating that the results have already converged with respect to $m_\text{max}$. The dashed lines are linear fits, showing that the deviations follow a power law with an exponent close to 1.}

\subsection{Vertex operator expectation values}

\begin{figure*}[t!]
    \centering
    \includegraphics[width=\linewidth]{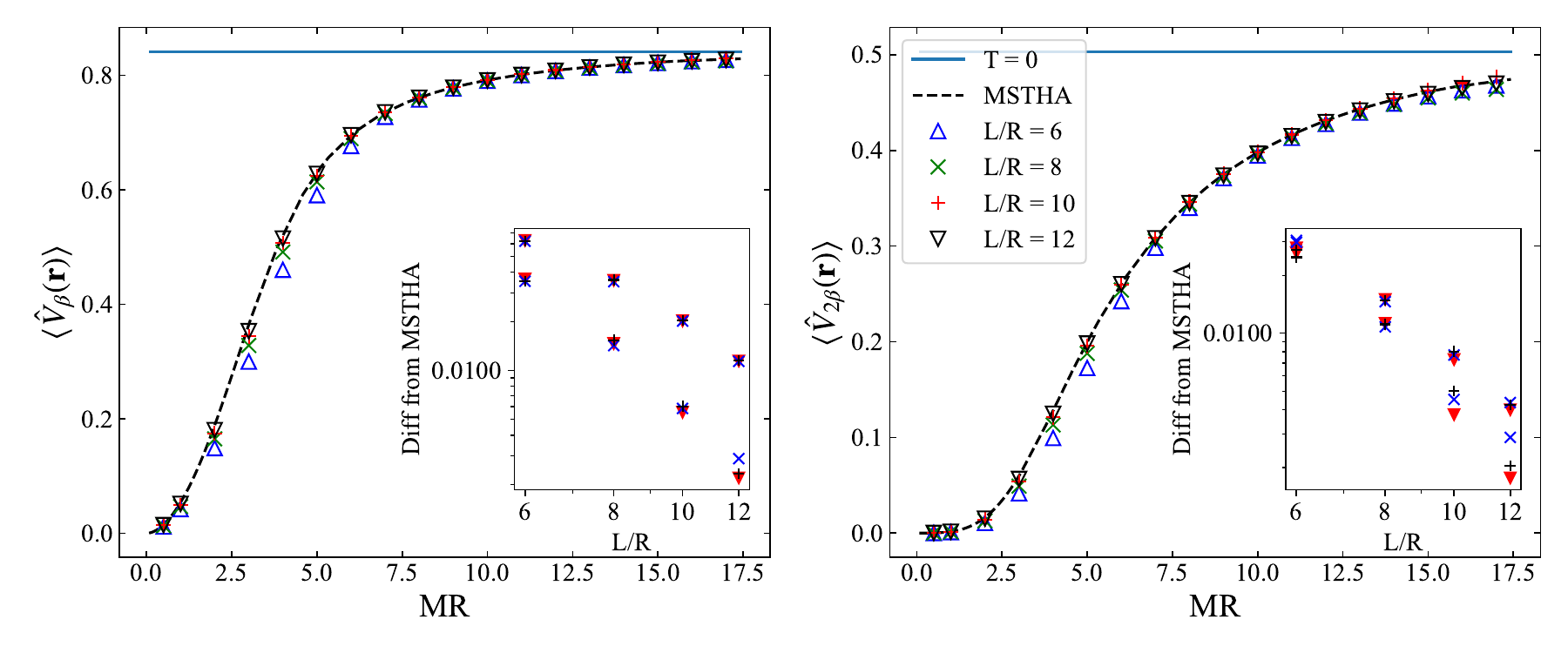}
     \caption{Expectation value of the $\hat V_{\beta}(\br)$ (left) and $\hat V_{2\beta}(\br)$ (right) vertex operator for $\beta^2=8\pi/25$ ($\Delta=2/25$). \emph{Main panels:} Expectation values as functions of the dimensionless temperature $MR$. The MRS results obtained with mode cutoff $m_\text{max}=20$ for different $L/R$ ratios are shown in symbols, while the dashed line represents the MSTHA data. The horizontal solid blue lines show the exact $T=0$ result. \emph{Insets:} Difference between the MRS and the MSTHA results at $MR = 2.9$ and $MR = 5.3$ on the left and at $MR = 3.82$ and $MR = 5.3$ on the right, using different mode cutoffs ($m_\text{max}=20$ in red triangles, 25 in blue crosses, 30 in black plus signs). The MRS results were obtained using $\sim10^6$ Monte Carlo samples, the statistical errors are smaller than the symbol size.
     }
     \label{fig:k008V}
\end{figure*}

\begin{figure*}[t!]
    \centering
    \includegraphics[width=\linewidth]{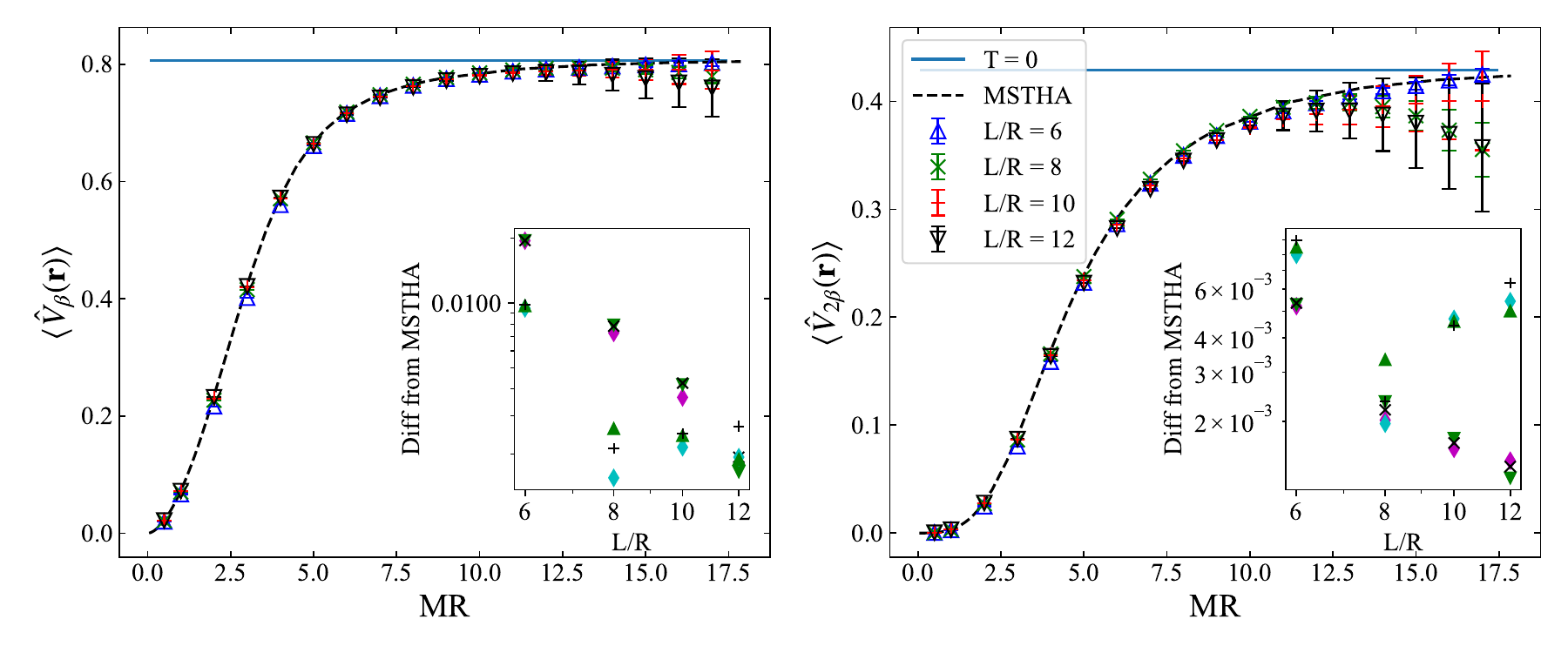}
     \caption{Expectation value of the $\hat V_{\beta}(\br)$ (left) and $\hat V_{2\beta}(\br)$ (right) vertex operator for $\beta^2=8\pi/15$ ($\Delta=2/15$). \emph{Main panels:} expectation values as functions of the dimensionless temperature $MR$. The MRS results obtained with mode cutoff $m_\text{max}=20$ for different $L/R$ ratios are shown in symbols, while the dashed line represents the MSTHA data. The horizontal solid blue lines show the exact $T=0$ result. In the MRS $N_\text{s}=5\cdot10^5$ Monte Carlo samples were used, the error bars show the standard deviation of the sampling, increasing whenever $MR$ or the ratio $L/R$ becomes larger. In particular, the observed deviation at low temperatures (large $MR$) signals the need for more MC samples to get accurate results. \emph{Insets:} Difference between the MRS and the MSTHA results at fixed values of $MR = 3$ for different numbers of Monte Carlo samples (red triangle $N_\text{s}=3\cdot10^5$, black plus $4\cdot10^5$, blue cross $5\cdot10^5$). 
     }
     \label{fig:k0133V}
\end{figure*}

Despite the integrability of the sine--Gordon model, there is no analytic result for its finite temperature expectation values. The only exact result is for zero temperature \cite{Lukyanov1996}, which we quote in  Appendix \ref{app:vev}. At nonzero temperature, we benchmark the numerical results of the MRS approach with results obtained using the recently developed MSTHA method \cite{2024PhRvB.109a4308S}. Exploiting the Lorentz invariance of the sine--Gordon theory, expectation values calculated at zero temperature on a ring of circumference $R$ are equal to finite temperature expectation values at inverse temperature $R$ in infinite volume. The accuracy of the truncated Hamiltonian approach can be well controlled by analysing the convergence of the results with respect to increasing the energy cutoff. In the following, the truncation parameter is chosen so that the results are accurate up to (at least) four decimal places.

We show the results of the two methods for the 1-point functions $\expv{\hat V_\beta(\br)}$ and $\expv{\hat V_{2\beta}(\br)}$, together with the analytic $T=0$ result, in Fig.\ \ref{fig:k008V} for $\beta^2=8\pi/25$ ($\Delta=2/25$) and in Fig.\ \ref{fig:k0133V} for $\beta^2=8\pi/15$ ($\Delta=2/15$). To track the finite size effects in the MRS method, we used different $L/R$ ratios and calculated the expectation value at the origin (at the centre of the cylinder). At small and medium inverse temperatures, the deviations are due to finite size effects, and by increasing the $L/R$ ratio, the agreement converges as a power law in $L/R$. This is also demonstrated in the insets of both figures, where we plot the difference between the MSTHA and the MRS results. In the inset of Fig.\ \ref{fig:k008V}, these differences are shown for two different $MR$ values and for three different mode cutoffs (see the caption for details), demonstrating that the results are nicely converged in the Fourier mode cutoff. 
For large $MR$ (low temperatures), we find deviations at larger $L/R$ due to an insufficient number of Monte Carlo samples that become more severe as we increase $L/R$. This is natural as larger $L/R$ requires more modes, and more modes require more averaging.
We examine this further in Fig.\ \ref{fig:k0133V}, where the convergence in $L/R$ is no longer monotonic at sufficiently low temperatures.
In the insets of Fig.\ \ref{fig:k0133V}, we examine the dependence on the number of Monte Carlo samples $N_\text{s}$ of the results, finding that our estimates are nicely converged at small $L/R$ ratios, but this increases the number of required Monte Carlo samples.

The expectation value of $\expv{\hat V_\beta(\br)}$ can also be computed from the free energy via the Hellmann--Feynman theorem \cite{hellman1937einfuhrung,1939PhRv...56..340F}, making it accessible to the NLIE approach as well; {we note that the result obtained from the NLIE agrees with the MSTHA to a high accuracy}. This is not the case for the other vertex operators for which the only viable method is the MSTHA, which we use to verify the MRS results in the right panels of Figs. \ref{fig:k008V} and \ref{fig:k0133V}.

\section{Conclusions and Outlook}\label{sec:conclusions}

We generalized the method of random surfaces to compute one-point functions of exponential operators in the sine--Gordon quantum field theory at finite temperature. We demonstrated the simplicity and effectiveness of the method of random surfaces and validated its convergence by comparing the free energy and the one-point functions to the results of alternative nonperturbative approaches, such as the exact NLIE relying on integrability and the numerical MSTHA truncated Hamiltonian method, finding excellent agreement.

This work constitutes the first step in developing the method. The next step is the computation of finite-temperature multi-point correlation functions, which will be reported soon in another publication \cite{mrs_correlators}, where such comparisons are no longer possible. The method is clearly versatile enough to treat additional refinements of the model, such as spatially inhomogeneous couplings and the inclusion of a disorder (notably being able to continue to work when direct quantum Monte Carlo treatments suffer from a sign problem). It is capable of treating finite systems with boundary conditions. Additionally, the method of random surfaces has the potential to be extended to time evolution after a quantum quench,  providing a promising alternative to the existing semi-classical approximations \cite{2013PhRvL.110i0404D,2016PhRvE..93f2101K,2017PhRvL.119j0603M,2019JSMTE..08.4012V,2019PhRvA.100a3613H} and Hamiltonian truncation methods \cite{2018PhRvL.121k0402K,2019PhRvA.100a3613H,2024PhRvB.109a4308S}. These features make the MRS an invaluable tool for the description of experimental realizations of sine-Gordon quantum field theory \cite{2010PhRvL.105s0403C,2010Natur.466..597H,2017Natur.545..323S,PRXQuantum.4.030308,2021NuPhB.96815445R,Wybo2022} and we expect that its applicability will be even further reaching than previously anticipated. 

\begin{acknowledgments}
We thank Adilet Imambakov for inspiring this project many years ago and for many insightful discussions.
This work was supported by the National Research, Development and Innovation Office (NKFIH) through the OTKA Grants ANN 142584 and K 138606. DSz was also partially supported by the ÚNKP-23-3-II-BME-182 New National Excellence Program of the Ministry for Culture and Innovation from the source of the National Research, Development and Innovation Fund, while GT was also partially supported by the NKFIH grant ``Quantum Information National Laboratory'' (Grant No. 2022-2.1.1-NL-2022-00004). 
J.H.P. is partially supported by NSF Career Grant No.~DMR-1941569. This work was performed in part at the Aspen Center for Physics, which is supported by National Science Foundation grant PHY-2210452 (J.H.P.).
\end{acknowledgments}

\bibliographystyle{utphys}
\bibliography{sG_rand_surf}

\clearpage

\appendix

\onecolumngrid

\section{Fourier modes}
\label{app:modes}

\begin{figure*}[t!]
    \centering
    \includegraphics[width=0.49\linewidth]{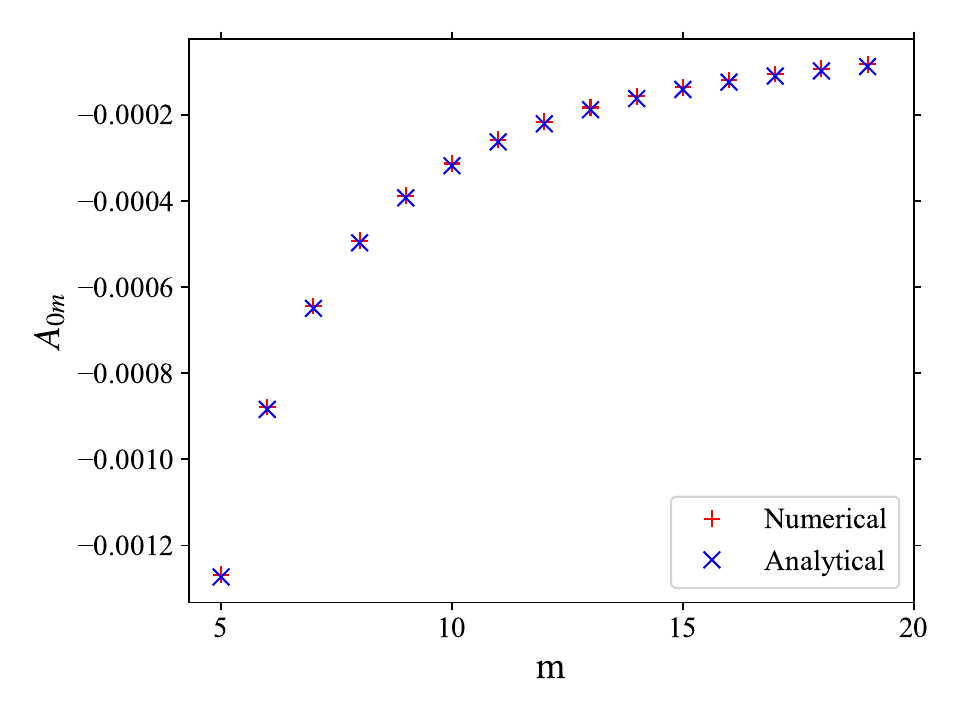}
    \hfill
    \includegraphics[width=0.49\linewidth]{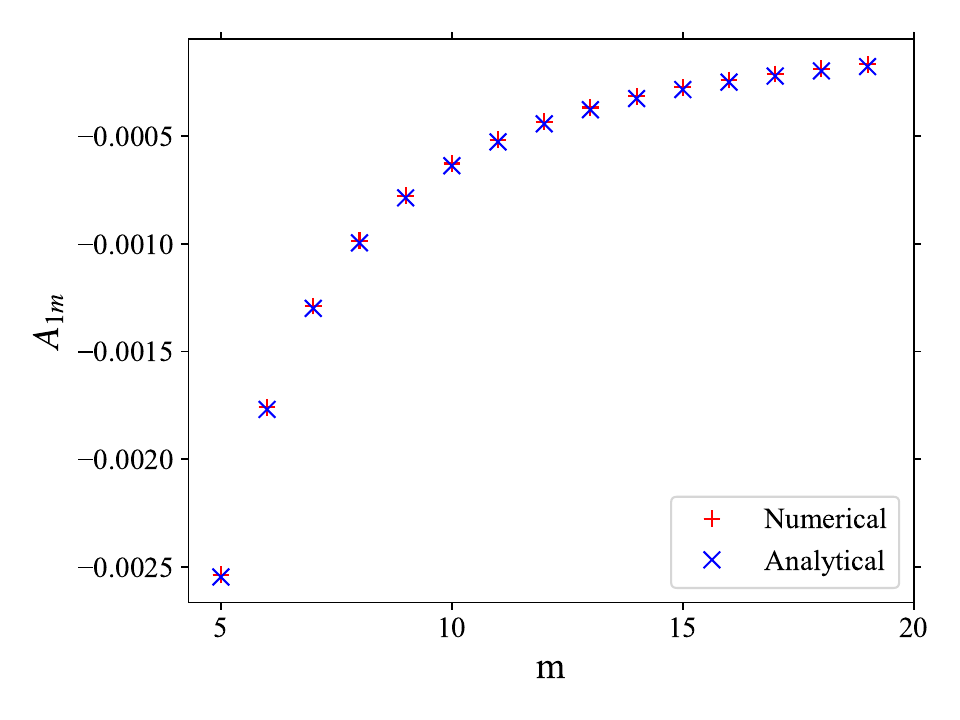}
    \caption{Numerical and analytical values for the Fourier modes $A_{0m}$ and $A_{1m}$ for $L/R=10$.}
    \label{fig:nmmodes1}
\end{figure*}

\begin{figure*}[t!]
    \centering
    \includegraphics[width=0.49\linewidth]{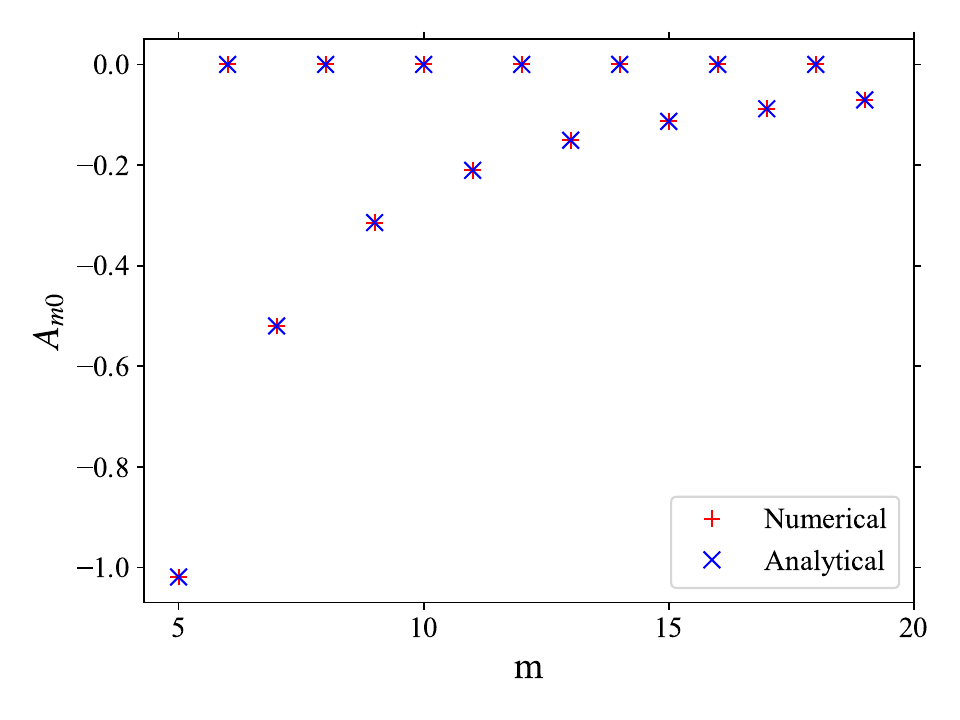}
    \hfill
    \includegraphics[width=0.49\linewidth]{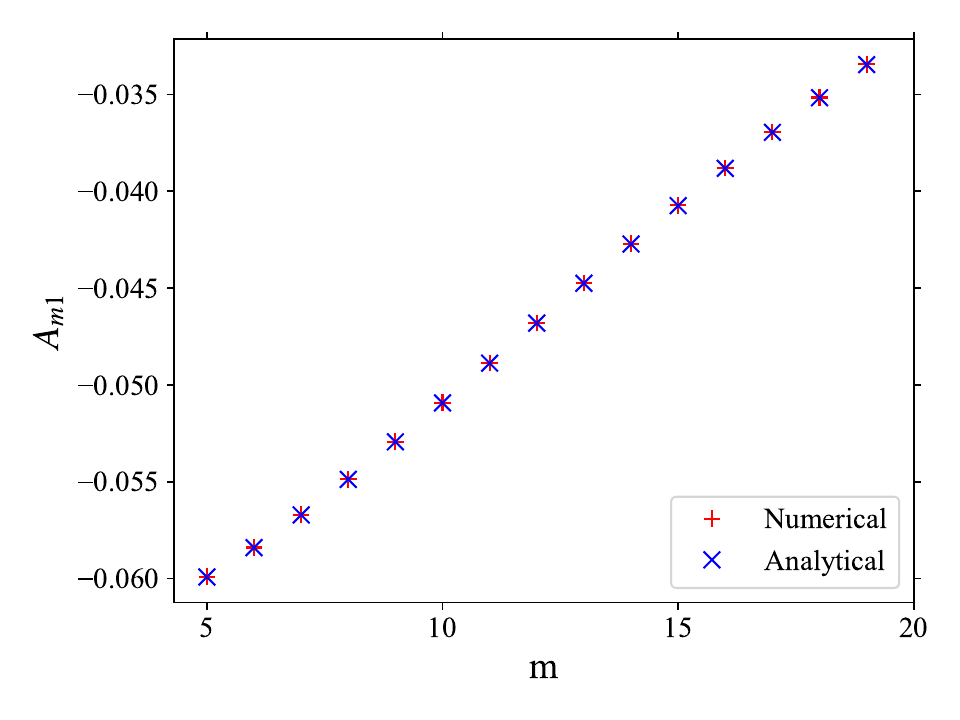}
    \caption{Numerical and analytical values for the Fourier modes $A_{m0}$ and $A_{m1}$ for $L/R=10$. 
    }
    \label{fig:nmmodes2}
\end{figure*}

In this appendix, we analyze the Fourier coefficients appearing in Eqs. \eqref{eq:Amn} in the main text. Let us recall that we consider the translational invariant Green's function of the infinite cylinder
\begin{align}
\label{eq:Greens1}
G(\br, 0) 
= -\log \left|\frac1{\pi}\sinh\left(\frac{\pi}{R}(x+i\tau)\right)\right|^2
\end{align}
with $\br=(x,\tau)$ on the Euclidean space-time domain $x\in[- L, L]$, $\tau\in[-R,R]$. It is periodic in the $\tau$-direction with period $R$, and we periodically extend it in the $x$-direction with period $2 L$. Therefore, it can be expanded in a double Fourier series: 
\begin{align}
G(\br_,0)= \sum_{m,n=0}^{\infty}B_{mn}\cos(\pi mx/ L)\cos(2\pi n \tau/R)\,.
\end{align}
We observe that the Fourier coefficients $B_{mn}$ are numerically very close to the simple expressions
\begin{subequations}
\label{eq:Amnapp}
\begin{align}
  A_{00} = 2\log(2\pi) - \pi L/R\,,
\end{align}
\begin{align}
  A_{m0} = \begin{cases}
    \frac{8L}{\pi R} \frac1{m^2} & \quad\text{m odd,} \\
    0 & \quad\text{m even, } m\neq0\,,
\end{cases}
\end{align}
and
\begin{align}
    A_{mn} = \frac{1}{1 + \delta_{m,0}} \frac{8\pi}{LR} \frac{1}{\left(\frac{\pi m}{L} \right)^2 + \left(\frac{2\pi n}{R} \right)^2}\,,\qquad n\neq0\,.
\end{align}
\end{subequations}
This is shown for $L/R=10$ in Fig. \ref{fig:nmmodes1} for the coefficients $A_{0m}$, $A_{1m}$ and in Fig. \ref{fig:nmmodes2} for $A_{m0}$ and $A_{m1}$. Increasing the accuracy of the numerical integration decreases the difference between the analytical and numerical values, showing that the analytical expressions are indeed correct.

To understand these coefficients better, we take the Laplacian of the function $\bar G(x,\tau)$ defined by the analytic coefficients. The $(n,m)$-dependence of the coefficients cancels with the factors coming from the derivatives of the cosine functions, and we are left with
\begin{multline}
    \Delta \bar G(x,\tau) = \Delta \sum_{m,n=0}^{\infty}A_{mn}\cos(\pi mx/ L)\cos(2\pi n \tau/R) \\
    = 
    \frac{4\pi}{R L} \left[2\sum_{\substack{m=1\\m\text{ odd}}}^\infty\cos(\pi m x/L)
    + \sum_{n=1}^\infty\cos(2\pi n \tau/R) 
    + 2\sum_{n,m=1}^\infty\cos(\pi m x/L)\cos(2\pi n \tau/R)\right]\,.
\end{multline}
Applying the Poisson summation formula, this is equal to
\begin{multline}
\label{eq:Poisson}
    \frac{4\pi}{R L} \left[
    2\sum_{k\in\mathbb{Z}}(-1)^k\frac{L}2\delta(x-kL)
    +\sum_{\ell\in\mathbb{Z}}\frac{R}2\delta(\tau-\ell R)-\frac12
    +2\left(
    \sum_{k\in\mathbb{Z}}L\delta(x-2kL)-\frac12\right)
    \left(\sum_{\ell\in\mathbb{Z}}\frac{R}2\delta(\tau-\ell R)-\frac12
    \right)
    \right]\\
    =
     \frac{4\pi}{R L} \left[
     -L\sum_{k\text{ odd}}\delta(x-kL)+
     LR\sum_{k,\ell\in\mathbb{Z}}\delta(\tau-kR)\delta(x-2\ell L)
     \right]\,.
\end{multline}
On the domain $x\in\,]-L,L[\,$,
\begin{align}
    \Delta \bar G(x,\tau)
    = 4\pi \sum_{k\in\mathbb{Z}}\delta(\tau-kR)\delta(x)\,,
\end{align}
so it is the Green's function of the Laplace equation with periodic boundary condition in the $\tau$-direction. As $L\to\infty$, it becomes the Green's function on the infinite cylinder.

However, this does not specify the function completely, since it also depends on the boundary conditions. We now provide some analytic evidence supporting that $\bar G(x,\tau)$ goes to $G(\br,0)$ for $L\gg R$. First, consider our function at $x=0$,
\begin{align}
    \bar G(0,\tau) = \sum_{n=0}^\infty\left(\sum_{m=0}^\infty A_{mn}\right) \cos(2\pi n \tau/R) = 
    \sum_{n=0}^\infty \frac2n \coth\left(\frac{2\pi n L}{R}\right)\cos(2\pi n \tau/R)\,.
\end{align}
For $L\gg R$, the coefficients are exponentially close to $2/n$, and we obtain
\begin{align}
    \bar G(0,\tau) \approx \sum_{n=0}^\infty \frac2n \cos(2\pi n \tau/R) = -2\log\left[\frac1\pi \sin\left(\frac{\pi\tau}R\right)\right] = G(\br,0)
\end{align}
for $\br=(0,\tau)$.

As a second check, let us evaluate the function at $x=\pm L$:
\begin{align}
    \bar G(\pm L,\tau) = \sum_{n=0}^\infty\left(\sum_{m=0}^\infty A_{mn}(-1)^m\right) \cos(2\pi n \tau/R) = 
    2\log(2\pi) -\frac{2\pi L}R + \sum_{n=1}^\infty \frac2{n \sinh\left(\frac{2\pi n L}{R}\right)}\cos(2\pi n \tau/R)\,.
\end{align}
For $L\gg R$, the coefficients in the sum are exponentially suppressed and we are left with the first two terms. It is easy to check that for $\br=(\pm L,\tau)$
\begin{align}
    G(\br,0)\approx -\log \left|\frac{e^{\pi(L+i\tau)/R}}{2\pi}\right|^2 = 2\log(2\pi) -\frac{2\pi L}R\,,
\end{align}
up to exponential precision in $2\pi L/R$, so $\bar G(\pm L,\tau)$ rapidly approaches $G(\br,0)$.

\begin{figure}
    \centering
    \includegraphics[width=0.49\linewidth]{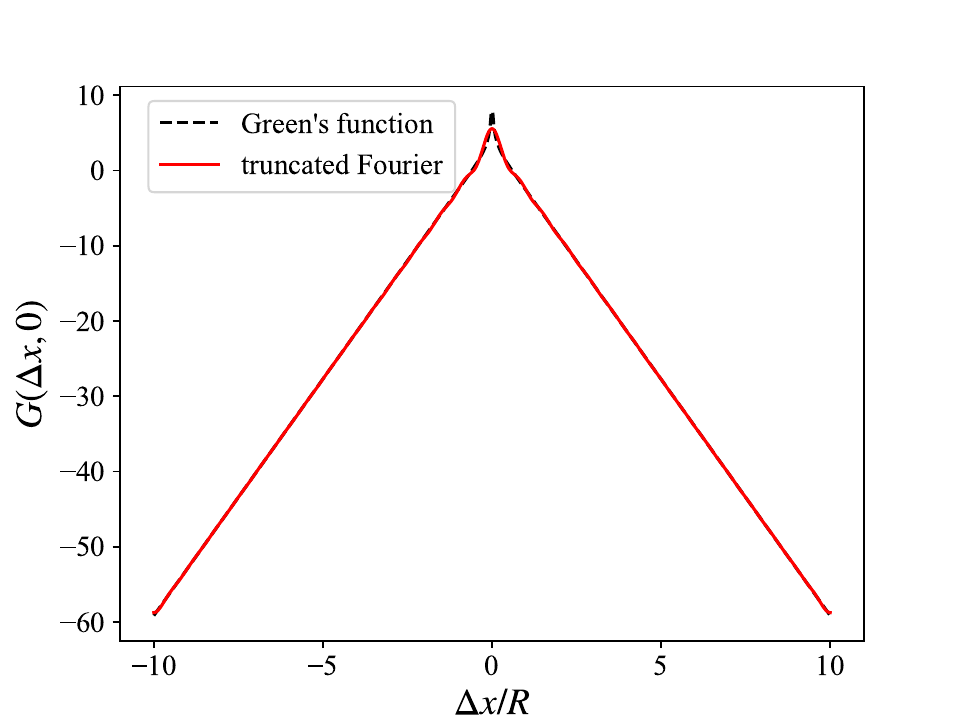}
    \hfill
    \includegraphics[width=0.49\linewidth]{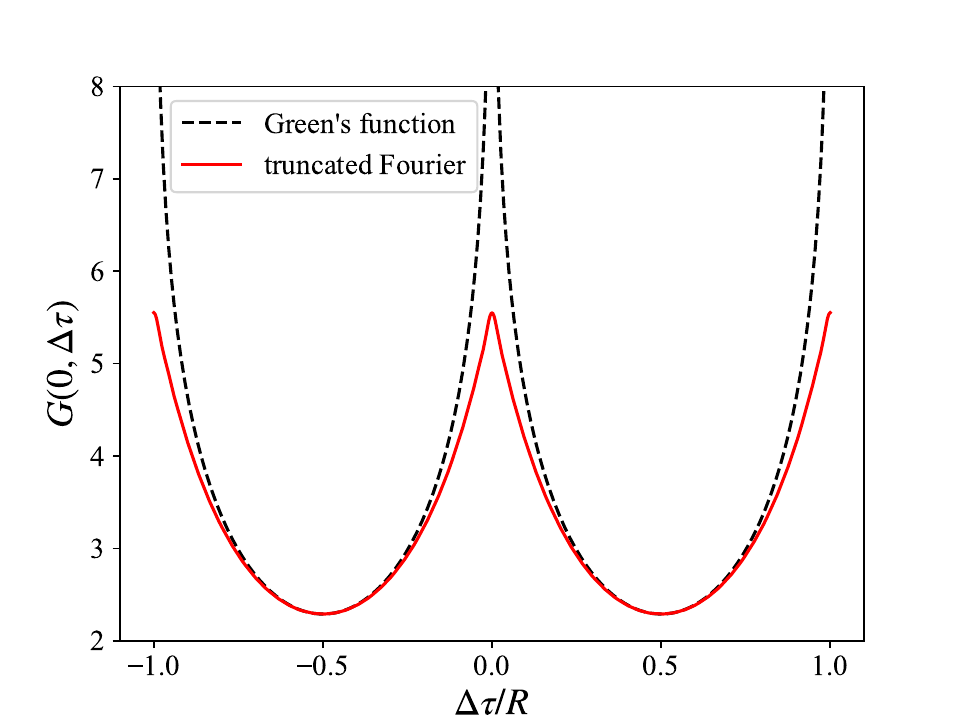}
    \caption{Sections of the exact (black dashed line) and the mode truncated (solid red line) Green's function $G(\br,0)$ at $\br=(x,0)$ (left) and $\br=(0,\tau)$ (right) for $m_\text{max}=30$ and $L/R=10$.}
    \label{fig:Gtrunc}
\end{figure}

{These analytic derivations hold for infinitely many modes. In the numerical calculations, the double Fourier series is truncated at a mode number $m_\text{max}$, which causes deviations from the exact Green's function even at large enough $L/R$. We show this truncation effect in Fig.\ \ref{fig:Gtrunc} where we compare $\tau=0$ and $x=0$ sections of the Green's function with the truncated Fourier series approximation. Naturally, there are significant deviations near the divergence at the origin. Nevertheless, the fact that it is an integrable singularity makes it possible to obtain accurate results.}

\section{The nonlinear integral equation}\label{app:nlie}
The nonlinear integral equation (NLIE) \cite{1991JPhA...24.3111K,1995NuPhB.438..413D} enables the calculation of the exact free energy density of sine--Gordon theory in an infinite volume at finite temperature $T$. It has the form 
\begin{align}
Z(\theta)=&\frac{M}{T} \sinh \theta - i\int\limits_{-\infty}^\infty \text{d}\theta' G(\theta-\theta'-i\varepsilon)\log\left( 1+ e^{i Z(\theta'+i\varepsilon)}\right)
+ i\int\limits_{-\infty}^\infty \text{d}\theta' G(\theta-\theta'+i\varepsilon)\log\left( 1+ e^{-i Z(\theta'-i\varepsilon)}\right)\,,\nonumber\\
&G(\theta)=\frac{1}{2\pi}
\int\limits_{-\infty}^{\infty} \text{d}t
\frac{\sinh\left(\frac{t\pi}{2}(\xi-1)\right)}{2\sinh\left(\frac{\pi\xi t}{2}\right)\cosh\left(\frac{\pi t}{2}\right)}{e}^{\mathrm{i}\theta t}\,,
\end{align}    
which can be solved iteratively for the function $Z(\theta)$, from which the free energy density $\tilde{f}$ defined in Eq.~\eqref{eq:free_energy_definition} can be obtained as
\begin{align}
    \frac{\tilde{f}}{T}=-2\,\mathrm{Im}\int\limits_{-\infty}^\infty \frac{\mathrm{d}\theta}{2\pi} M\sinh(\theta+i\varepsilon)\log\left( 1+ e^{i Z(\theta+i\varepsilon)}\right)\,.
    \label{eq:nlie_free_en}
\end{align}

\section{Vertex operator expectation values at zero temperature}
\label{app:vev}

The exact expectation values of vertex operators at zero temperature expectation were obtained by Lukyanov and Zamolodchikov \cite{Lukyanov1996} with the result:
\begin{align}
    \expv{\hat V_\alpha}  = \left[\frac{m \Gamma\left(\frac{1+\xi}{2}\right)\Gamma\left(1-\frac{\xi}{2}\right)}{4\sqrt{\pi}} \right]^{\Delta_\alpha} 
    \exp{\int_0^{\infty}\frac{dt}{t}\left[
    \frac{\sinh^2(2\alpha\beta t)}{2\sinh(\beta^2 t)\sinh(8\pi t)\cosh((8\pi-\beta^2)t)} - \frac{\alpha^2}{4\pi} e^{-16\pi t}\right]}\,,
\end{align}
where $\Delta_{\alpha} = \alpha^2/(4\pi)$, $m = 2M\sin(\pi\xi/2)$ and $\xi = \beta^2/(8\pi - \beta^2)$.

\section{Additional numerical results}
\label{app:extraplots}

\subsection{Numerical integration via the trapezoidal rule}
\label{app:discr}

\begin{figure}[t!]
    \centering
    \includegraphics[width=0.5\textwidth]{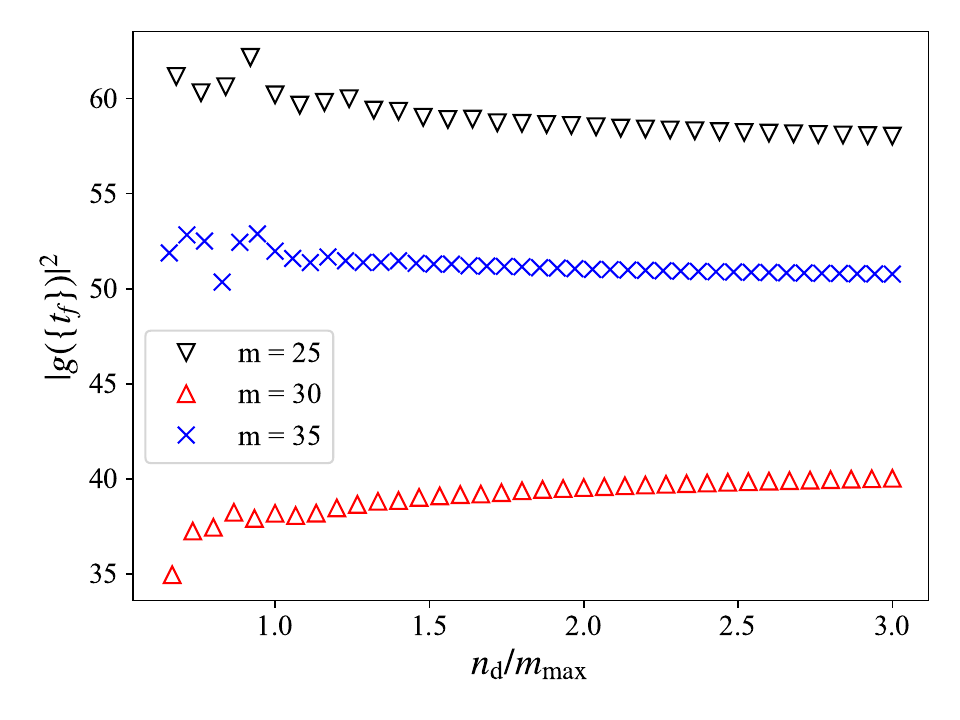}
    \caption{\label{fig:discr} Distinct typical instances (i.e. different samples of random $t_\ff$) of $\left| g(\{t_\ff\}) \right|^2$ at three different $m_\text{max}$ maximal mode numbers versus its ratio with the number of grid points $n_\text{d}$ along each direction,  that is used in the numerical integration. 
    ($\beta^2=2\pi/5$, $L/R=6$.)}
\end{figure}

As discussed in the main text, we compute the integrals 
\begin{align}
    g(\{t_\ff\}) = C\int d\bu \exp\left\{ \sum_\ff^{m_{\mathrm{max}}}  it_\ff\sqrt{\Delta G_\ff}\,\psi_\ff(\bu) 
    \right\}
\end{align}
by discretizing the mode functions $\psi_\ff(\bu)$ and calculating the Riemann sum. The discretization must be fine enough to get a sufficiently accurate result. {In Fig.\ \ref{fig:discr}, we plot the value of the integral for three different mode cutoffs $m_\text{max}$ versus the ratio of the number of discretization points $n_\text{d}$ (used both in the $x$ and $\tau$ directions) and the mode cutoff $m_\text{max}$. 
A similar accuracy is achieved for the same ratio in the different cases, showing that the necessary resolution of the grid (i.e. $n_\text{d}$) for a desired accuracy is proportional to $m_\text{max}$. In our simulations, we used  $n_\text{d}/m_\text{max}$ between 2.5 and 4.}
\subsection{Position dependence of the 1-point function}
\label{app:posdependence}

\begin{figure*}[t!]
    \centering
    \includegraphics[width=0.5\linewidth]{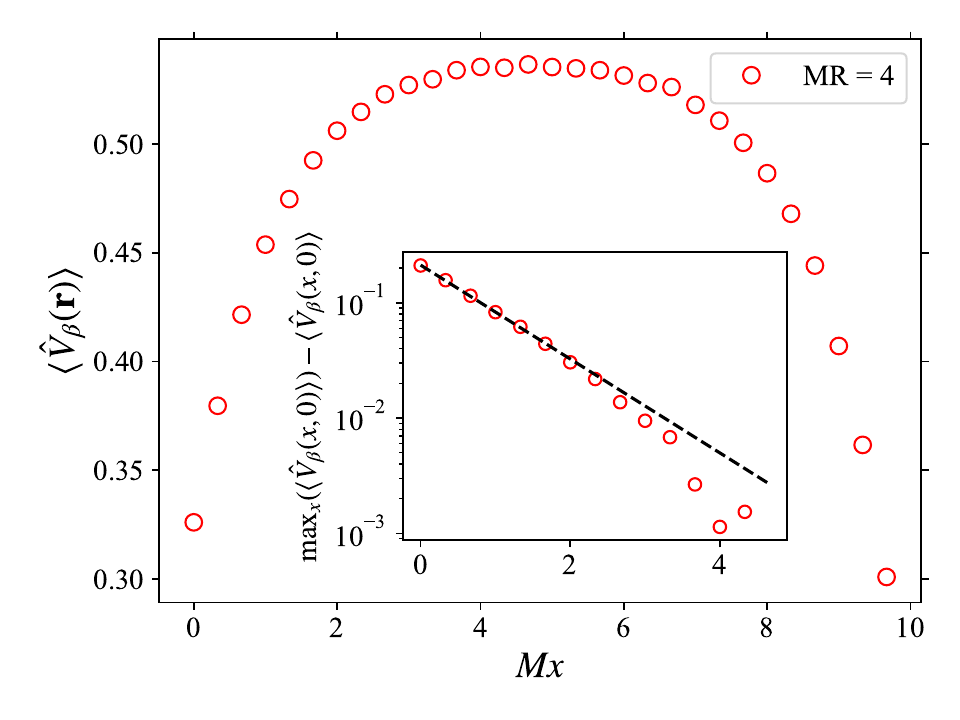}
    \caption{\label{fig:xdependence} Position dependence of the expectation value for $\beta^2=2\pi/5$, $L/R=10$, $MR=4$, and $m_\text{max}=30$. The inset shows the deviation from the maximal value at the centre on a log-lin scale, showing exponential convergence with a correlation length $0.9374/M$.}
\end{figure*}

{Since we integrate the multipoint functions over a finite space-time domain, our numerical results for the vertex operator expectation values are not translationally invariant in the $x$-direction. We plot the $x$-dependence of the expectation value $\expv{V_\beta(x,0}$ in Fig.\ \ref{fig:xdependence}. The inset shows that the deviation decays exponentially with the distance from the edges. The associated correlation length depends on the dimensionless inverse temperature $MR$.}

\twocolumngrid

\end{document}